# Fiber-integrated NV Magnetometer with Microcontroller-based Software Lock-in Technique

*Qilong Wu[1†], Xuan-Ming Shen[1†], Yuan Zhang[1, 2 *], Ying-Geng Shan[1], Hui-Hui Yu[1], Jing-Hao Zhang[1], Jiahui Chen[1], Yan Wang[3*], Xun Yang[1], Yong-Zhi Tian[1], Lijun Wang[4], and Chong-Xin Shan[1*]*

[1] Henan Key Laboratory of Diamond Materials and Devices, School of Physics and Laboratory of Zhongyuan Light, Zhengzhou University, Zhengzhou 450052, China

[2] Institute of Quantum Materials and Physics, Henan Academy of Sciences, Zhengzhou 450046, China

[3] School of Electronics and Information, Academy for Quantum Science and Technology, Zhengzhou University of Light Industry, Zhengzhou 450001, China

[4] State Key Laboratory of Luminescence and Applications, Changchun Institute of Optics, Fine Mechanics and Physics, Chinese Academy of Sciences, Changchun 130033, China

*Email: yzhuaudipc@zzu.edu.cn; ywang@zzuli.edu.cn; cxshan@zzu.edu.cn

**Abstract:**

Fiber-integrated nitrogen-vacancy (NV) magnetometers possess high sensitivity, integration, and flexibility, and thus have been explored extensively for industrial applications. While most studies have focused on the optimization of the quantum sensing head, less attention has been paid to the frequently employed professional, expensive, and bulky electronics, which hinder their practical applications. In this article, we fabricate a fiber-integrated NV magnetometer and develop a low-cost microcontroller-based software lock-in technique. In this technique, a microcontroller coordinates efficiently a microwave source chip and an analog-to-digital converter, and a program mimicking the lock-in mechanism realizes microwave frequency-modulated optically detected magnetic resonance of NV centers. As a result, with our setup and technique, we have realized the detection of weak magnetic field with a sensitivity of 93 nT/$\sqrt{\text{Hz}}$, which is comparable to what obtained with bulky and professional devices. Furthermore, we demonstrated real-time magnetic field detection, achieving a standard deviation of 488 nT. Our work provides a novel and cost-effective technique for electronic miniaturization, thereby potentially accelerating the industrial application of NV magnetometers.

# 1. Introduction

Nitrogen-vacancy (NV) center is formed in diamond lattice with a substitutional nitrogen atom adjacent to a vacancy.[1-3] Because of the unique electronic and spin structure, the NV center spin can be polarized with laser illumination, read-out through fluorescence, and manipulated with microwave radiation. More importantly, owing to the relatively clean spin environment in diamond, the NV center spin has relatively long quantum coherence time even at room temperature.[4,5] The above features make NV center an ideal system for quantum information[6-8] and quantum sensing.[1,2,9] Quantum sensing with NV centers typically explores the coupling of the ground state spin to magnetic fields via Zeeman effect,[10-12] electric fields through electronic dipole moment,[13,14] temperature via lattice constant-dependent zero-field splitting (ZFS)[15,16] and stress or pressure via strain-dependent ZFS.[17-19]

To achieve flexibility, stability, and simplicity, as required in industrial applications, fiber-based NV magnetometers have been developed extensively in recent years.[20-23] In these devices, optical fibers are used to simplify optical excitation and fluorescence collection, while providing flexibility by separating the sensing head from the ancillary electronics. Continuous-wave optically detected magnetic resonance (ODMR) is usually preferred due to its simple sensing scheme involving only continuous-wave laser and microwave radiation. In addition, lock-in amplifier (LIA) has been explored to mitigate environmental noise and low-frequency electronic noise.[20] After continuously optimizing the sensing head over the past decade, researchers have pushed the magnetic sensitivity of the fiber-based NV magnetometer below 12 $pT/\sqrt{Hz}$.[24] The applications of such magnetometers have also been demonstrated, including high-voltage current monitoring,[25,26] charging detection of electric vehicles,[27,28] industrial non-destructive testing,[21,29] biomedical imaging, etc.[30,31]

Most studies on fiber-based NV magnetometers have focused on the optimization of the sensing head,[32-34] but less attention has been paid to the employed professional, expensive, and bulky electronics, which hinder their practical deployment. To address this problem, in this work, we fabricate a fiber-integrated NV magnetometer sensing head, and develop a low-cost chip-based software lock-in technique, in which a microcontroller chip coordinates efficiently a microwave source chip and an integrated analog-to-digital converter (ADC), and a program is designed to mimic the lock-in mechanism to realize microwave frequency-modulated (FM) ODMR. With our setup and technique, we realize the FM-based LIA detection with arbitrary modulation depth, limited only by the bandwidth of the microwave source chip, and achieve a detection bandwidth of 40 Hz, which can be increased to 2 kHz by optimizing the frequency-switching speed. As a result, we realize the detection of weak magnetic field with a sensitivity down to $93 \text{ nT}/\sqrt{\text{Hz}}$, which is comparable with what achieved with an FPGA-based commercial LIA. Furthermore, we demonstrate a real-time detection of periodically modulated magnetic field, achieving a standard deviation of 488 nT. In the future, by further miniaturizing the electronics and optimizing the sensing head, we expect that a credit-card–sized board can be developed to power the fiber-based sensing head with a sub-nanotesla sensitivity,[35,36] which can potentially accelerate the applications of NV magnetometers.

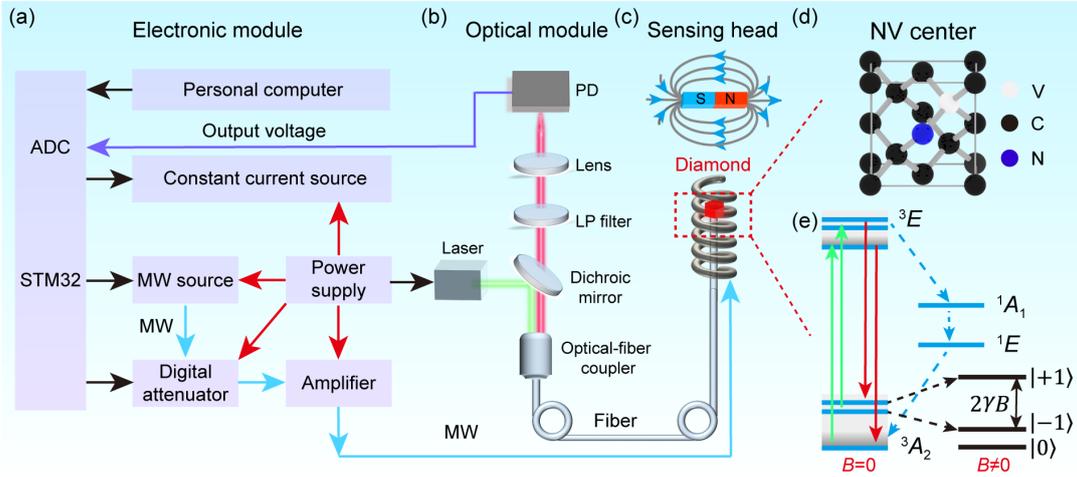

**Figure 1. Schematics of the fiber-based NV magnetometer.** a-c) Electronic module, optical module, and sensing head, respectively. d-e) Atomic structure and energy level diagram of the NV center. More details are provided in the main text.

## 2. Fiber-based NV Magnetometer

Our fiber-based NV magnetometer comprises three main components: an electronic module, an optical module, and a sensing head, as illustrated in Figure 1. The electronic module (Figure 1a) employs a microcontroller (STM32F103ZET6) to control a microwave source chip (ADF4351) to provide the microwave radiation, which is subsequently attenuated by a digital attenuator (HMC624A) and amplified by a power amplifier (KDT0308B-004). In addition, we utilize a voltage-controlled constant current source (KW-VCCS1000) to drive a laser diode, and use the ADC integrated in the microcontroller to convert the output voltage from a photodetector (PD, Thorlabs, PDA36A2). Photographs and circuit connections of the electronic module are provided in sect. S1 of the Supporting Information (SI).

The optical module is shown in Figure 1b. A 520 nm laser beam with power up to 130 mW is reflected by a dichroic mirror (OFD1LP-650) and coupled into a multi-mode fiber with a 105 μm core diameter through an optical-fiber coupler (FAC2-532-PC) with 90% efficiency. The fluorescence emitted from a 40 μm diamond with 3.5 ppm NV centers is collected by the same fiber, filtered by a 650 nm long-pass (LP) filter (OFE1LP-650), and subsequently focused onto the PD using a convex lens (OLC220136). All optical components are mounted on a 3D-printed optical board, providing a cost-effective platform for rapid prototyping and iterative optimization, which can be further optimized as an integrated form in the future (see sect. S1 of SI).[37] The sensing head is shown in Figure 1c. The microdiamond is attached to the tip of the optical fiber using UV-curable adhesive, and is wrapped with a 150 μm diameter copper wire (as a microwave antenna). For more details, see sect. S2 of SI. In addition, a bias magnetic field is applied to the NV centers with a magnet below the sensing head.

Before presenting the experimental results, we briefly summarize the magnetic field sensing principle of NV centers. The atomic structure and energy level diagram of NV center are shown in Figures 1d and 1e, respectively. Under a green laser excitation, the NV center is excited to the triplet excited state $^3E$, and then relaxes back to the triplet ground state $^3A_2$ through either the spin-preserving radiative process or the spin-selective non-radiative process through singlet levels $^1A_1$ and $^1E$. Because the NV center spin flips from the levels $m_s = \pm 1$ to $m_s = 0$ during the latter process, the ground state spin can be effectively polarized to the latter spin level under continuous laser illumination, and the NV center on the level $m_s = 0$ emits also more photons than that on the levels $m_s = \pm 1$. By applying microwave radiation to the NV centers, resonant microwave absorption changes the NV spin polarization, and leads to a change of the photoluminescence (PL).[2] The ODMR spectrum is achieved by recording the fluorescence change as a function of the microwave frequency (Figure 2a). When an external magnetic field $B$ is applied, the spin levels $m_s = -1$ and $m_s = +1$ shift downward and upward, respectively, by $\gamma B$ (with the gyromagnetic ratio $\gamma = 2.8$ MHz/Gauss). Consequently, the ODMR splits into two dips with a frequency separation $\delta f = 2\gamma B$, from which the magnetic field can be extracted.[1] Here, the outermost dips originate from the NV centers with axes aligned with the magnetic field, while the central dips arise from other NV centers (Figure 2b). In these measurements, the laser and microwave powers were about 80 mW and 25 dBm, respectively, and the average of 500 ADC-sampled voltages within 60 ms was used to reflect the fluorescence intensity of NV centers.

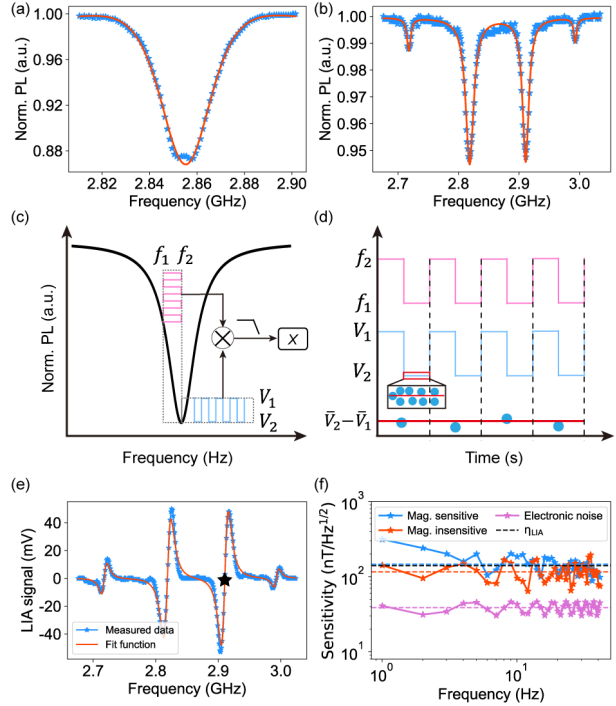

**Figure 2. ODMR spectra and software lock-in mechanism.** a,b) ODMR spectra (blue stars) in the absence and presence of a magnetic field, respectively, which is aligned with one NV axis, and the fittings with Lorentzian function (red solid curves). c) Schematic of lock-in amplifier mechanism with a square-wave microwave frequency modulation (FM). d) Mimicked phase locking and filtering with the synchronized ADC sampling and averaging. e) FM ODMR spectrum (blue stars) and the fitting with the expression described in the main text (red line) for the modulation depth $f_m = 10$

MHz. f) Noise spectral density for the resonant microwave frequency (blue curve) as marked by black star in (e), for the off-resonant microwave frequency (red curve), for the electronic noise (green curve). The black dashed line shows the theoretically calculated magnetic field sensitivity, while the other dashed lines are the noise spectral density averaged over 40 Hz bandwidth.

### 3. Magnetic Sensing with Microcontroller-based Lock-in Technique

The principle of lock-in detection for FM ODMR is illustrated in Figure 2c. A periodic switching of the microwave frequencies leads to a periodic modulation of the NV fluorescence. Then, the reference signal and the fluorescence signal are multiplied, and the resulting signal is filtered by a low-pass filter to remove the high-frequency noise, which produces a constant output proportional to the amplitude of the fluorescence modulation. Inspired by this mechanism, we come up with a simple way to mimic it (Figure 2d). The microwave source chip is switched between two microwave frequencies $(f_1, f_2)$, where the center frequency $f_c = (f_1 + f_2)/2$ and the difference $f_m = |f_1 - f_2|$ correspond to the so-called carrier frequency and the modulation depth, respectively. Simultaneously, the ADC integrated in the microcontroller records two series of voltage from the PD $(\{V_1\}, \{V_2\})$, mimicking the phase locking mechanism. Then, we calculate the averaged voltage $(\bar{V}_1, \bar{V}_2)$ and their difference $V_{LIA} = \bar{V}_2 - \bar{V}_1$ to mimic the low pass filtering. In this method, by varying the voltage sampling time, the detection bandwidth can be varied, and is ultimately limited by the microwave frequency switching time. The partial code of our method is provided in sect. S3 of the SI.

The FM ODMR spectrum (Figure 2e) is obtained by recording the lock-in voltages as a function of the carrier frequency $f_c$ with a modulation depth $f_m = 10$ MHz. The FM ODMR spectrum is simply the derivative of the ODMR spectrum (Figure 2b) with respect to the microwave frequency, and thus the carrier frequencies leading to zero lock-in voltage are those resonant with the NV center spin transitions. Notably, the FM ODMR spectrum is consistent with that obtained with a system using commercial microwave sources and lock-in amplifier under the same experimental conditions (see sect. S4 of SI for more details). In the FM ODMR spectrum, the lock-in voltage near a resonance shows a linear dependence on the frequency $V_{LIA} = S(f - f_0)$ with the resonance frequency $f_0$ and the slope $S$, and thus the lock-in voltage $V_{LIA} = S\gamma(B - B_0)$ with the bias magnetic field $B_0$ can be used to extract the magnetic field $B$ in a fast way. To estimate the sensitivity of the magnetic field sensing, we fit the FM ODMR spectrum with the expression[38]

$$L(f) = \sum_{i=1}^{4} -\frac{32}{3\sqrt{3}} \frac{C_i v_i^3 (f - f_i)}{(4(f - f_i)^2 + v_i^2)^2}. \quad (1)$$

Here, for the $i$-th resonance, $C_i$ denotes the peak contrast, $f_i$ represents the resonance frequency, and $v_i$ corresponds to the ODMR linewidth. For each resonant feature, we can compute the linear slope $S_i = \frac{32\sqrt{3}}{9} \frac{C_i}{v_i}$ (e.g., $S_3$ = 16 mV/MHz for the third feature around 2.91 GHz). The magnetic field sensitivity can be theoretically estimated as[38]

$$\eta_{LIA} = \frac{\sigma}{\gamma S_i \sqrt{2 f_{ENBW}}}, \quad (2)$$

where $\sigma$ is the standard deviation of the LIA signal, and $f_{ENBW}$ is the equivalent noise bandwidth of the lock-in amplifier.

To determine the sensitivity of the magnetic field sensing under realistic conditions, we set the carrier frequency resonant to the third feature in Figure 2e, record a list of 415 lock-in voltages over 5 seconds. Then, we split the data into $\alpha = 1, \ldots, 5$ segments with 1 second each, calculate the power spectral density (PSD) using the Periodogram method,[39] and finally compute the noise spectral density $PSD/(\gamma S_i)$, see Figure 2f. The measured magnetic field sensitivity is about 313 $nT/\sqrt{Hz}$ at the 1 Hz frequency, and the average sensitivity over the entire frequency range is about 146 $nT/\sqrt{Hz}$ (blue curve). For a microwave frequency off-resonant to the NV centers, we estimate the sensitivity in the magnetic field insensitive case as 115 $nT/\sqrt{Hz}$ (red curve). If we switch off the laser and the microwave, we estimate the electronic noise-limited sensitivity as 38 $nT/\sqrt{Hz}$ (purple curve). Note that the sensitivity according to Equation (2) is estimated as $\eta_{LIA}$ = 142 $nT/\sqrt{Hz}$ with $\sigma$ = 0.57 mV and $f_{ENBW}$ =

40 Hz (black dashed line), which is consistent with the experimental results.

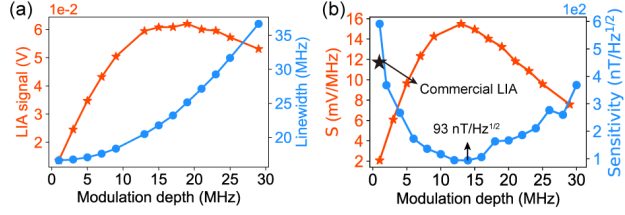

**Figure 3. Optimization of magnetic field sensitivity.** a) Maximal lock-in voltage $C_3$ (red curve, left axis) and the linewidth $v_3$ (blue curve, right axis) of the FM ODMR spectrum as a function of the frequency modulation depth $f_m$. b) Slope of linear fitting $S_3$ (red curve, left axis) and the magnetic field sensitivity (blue curve, right axis) as a function of $f_m$. Here, the black star indicates the sensitivity obtained with a commercial microwave source and lock-in amplifier under the same condition.

For a commercial microwave source (such as RIGOL, DSG 836A), the frequency modulation depth $f_m$ is usually limited, which might become a factor to constrain the achieved sensitivity. In contrast, in the software lock-in detection scheme as suggested here, $f_m$ can be significantly increased and is only limited by the bandwidth of the microwave source chip. Thus, in Figure 3, we optimize the magnetic field sensitivity by studying its dependence on $f_m$. Figure 3a shows that as $f_m$ increases from 1 MHz to 30 MHz, the maximal lock-in voltage $C_3$ increases first from 10 mV to the maximum of about 60 mV at $f_m = 18$ MHz and then decreases slightly (red curve, left axis). Meanwhile, the linewidth $v_3$ increases almost quadratically from 16 MHz to 37 MHz (blue curve, right axis). As a result, the zero-crossing slope $S_3$ approaches the maximum of 15.5 mV/MHz at $f_m = 13$ MHz, and the obtained sensitivity reduces from 591 nT/$\sqrt{\text{Hz}}$ at $f_m = 1$ MHz to 93 nT/$\sqrt{\text{Hz}}$ (Figure 3b). For a comparison, using a commercial microwave source and lock-in amplifier under the same condition, we achieve a sensitivity of 421 nT/$\sqrt{\text{Hz}}$ at $f_m = 1$ MHz (black star), which is only about 1.3 times better than our system. For more details, see sect. S4 of SI.

**4. Real-time Sensing of Magnetic Field**

After characterizing the sensitivity of our fiber-based NV magnetometer, we now demonstrate its application for real-time magnetic field sensing. To this end, we place an electromagnet below the sensing head, and modulate the applied magnetic field by changing its current. We adjusted the position of the sensing head to achieve the FM ODMR spectrum as shown in Figure 2e, and utilized the rightmost feature, i.e., the NV centers with axes parallel to the magnetic field (Figure 4a), in the following experiment. To mitigate the limitation due to the 12-bit ADC of the microcontroller, we have also amplified the PD voltage by 22 times with a TL082 chip before the ADC sampling. In this case, the linear fitting of the FM ODMR spectrum leads to a slope $S_4 = \overline{60 \text{ mV/MHz}}$ within a frequency range 15.3 MHz, and a sensitivity of 50 nT/$\sqrt{\text{Hz}}$ is achieved. We set the FM carrier frequency at the zero-crossing point and modulate the electromagnet current at 1 Hz. We observe the alternation of the lock-in voltage between 0.18 V and 0 V, which corresponds to a magnetic field variation between 0.2 mT and 0 mT (Figure 4b). To further evaluate the signal stability, we magnify the data for the last 1 second duration (Figure 4c), and extract a standard deviation of about 488 nT from the histogram analysis (Figure 4d), indicating the practical magnetic field resolution of our system.

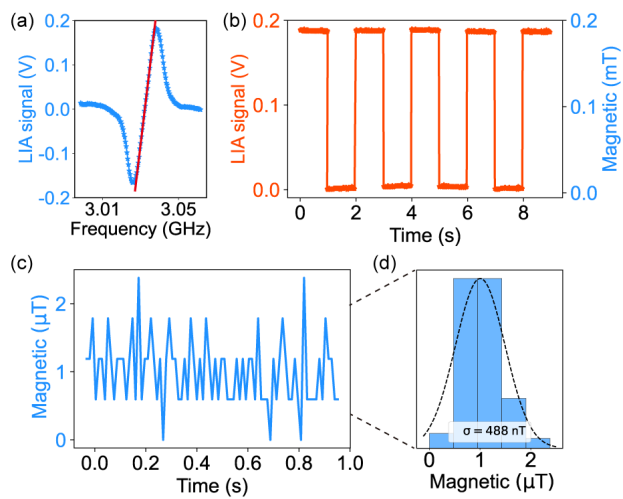

**Figure 4. Real-time magnetic field sensing.** a) An amplified spectrum of the rightmost FM ODMR in Figure 2e. The red line represents a linear fit. b) Continuous tracking of a magnetic field varying periodically at 1 Hz. c) Enlarged trace at the last 1 second duration. d) Corresponding histogram with a standard deviation of 488 nT.

## 5. Discussion and Conclusion

In summary, we fabricated a fiber-integrated NV magnetometer and developed a microcontroller-based software lock-in technique to realize frequency-modulated optically detected magnetic resonance. The system achieves a magnetic field sensitivity of $93\,\text{nT}/\sqrt{\text{Hz}}$, which is comparable to that with a commercial microwave source and lock-in amplifier. In addition, we demonstrate a real-time magnetic field sensing with a standard deviation of about 488 nT. In the future, our magnetometer can be further optimized. First, the 40 μm diamond can be replaced by a sample matching the fiber core size, and the sensor head can be coated with a silver layer[34] or use a micro-concave mirror to enhance fluorescence excitation and collection,[32] which can improve sensitivity by about one order of magnitude. Second, the laser diode can be replaced with a more powerful and stable laser, and a balanced photodetector can be used to mitigate the laser intensity noise. Third, the 12-bit ADC integrated in the microcontroller can be upgraded with a 16-bit or 24-bit ADC to increase the voltage resolution by 16 or 4096 times, which can reduce significantly the electronic noise. With all these improvements, we expect a magnetic field sensitivity down to sub-nanotesla level. In addition, we can miniaturize our electronics into a credit-card sized board, which will potentially accelerate the deployment of such magnetometers in industrial applications.

## Supporting Information

Supporting Information is available from the Wiley Online Library or from the author.

## Author Contributions

Q. Wu carried out the studies under the supervision of Y. Zhang, who devises the idea, the electronics, and the program. Y.-G. Shan constructed the electronics. X.-M. Shen, H.-H. Yu, J.-H. Zhang and J. C assisted with the optical setup. Y. Wang and X. Yang prepared the sensing head. Y.-Z. Tian helped with the programming. L. Wang and C.-X. Shan contributed to the manuscript writing.

**Acknowledgements**

This work was supported by the National Key R&D Program of China (Grant No. 2024YFE0105200), the National Natural Science Foundation of China (Grants No. 12422413 and No. 62475242), Science and Technology Major Project of Henan Province (231100230300), Henan Association for Science and Technology Youth Talent Support Program (2024HYTP024), Natural Science Foundation of Henan Province (252300421228), Central Plains Outstanding Young Talents Program.

**Conflict of interest**

The authors declare that they have no conflict of interest.

**Reference:**

[1] J. F. Barry, J. M. Schloss, E. Bauch, M. J. Turner, C. A. Hart, L. M. Pham, R. L. Walsworth, *Rev. Mod. Phys.* **2020**, *92*, 015004.

[2] J. Du, F. Shi, X. Kong, F. Jelezko, J. Wrachtrup, *Rev. Mod. Phys.* **2024**, *96*, 025001.

[3] G. Wolfowicz, F. J. Heremans, C. P. Anderson, S. Kanai, H. Seo, A. Gali, G. Galli, D. D. Awschalom, *Nat. Rev. Mate.* **2021**, *6*, 906.

[4] N. Bar-Gill, L. M. Pham, A. Jarmola, D. Budker, R. L. Walsworth, *Nat. Commun.* **2013**, *4*, 1743.

[5] S. Han, X. Ye, X. Zhou, Z. Liu, Y. Guo, M. Wang, W. Ji, Y. Wang, J. Du, *Sci. Adv.* **2025**, *11*, eadr9298.

[6] G. Waldherr, Y. Wang, S. Zaiser, M. Jamali, T. Schulte-Herbrüggen, H. Abe, T. Ohshima, J. Isoya, J. F. Du, P. Neumann, J. Wrachtrup, *Nature* **2014**, *506*, 204.

[7] T. H. Taminiau, J. Cramer, T. van der Sar, V. V. Dobrovitski, R. Hanson, *Nat. Nanotech.* **2014**, *9*, 171.

[8] N. Kalb, A. A. Reiserer, P. C. Humphreys, J. J. W. Bakermans, S. J. Kamerling, N. H. Nickerson, S. C. Benjamin, D. J. Twitchen, M. Markham, R. Hanson, *Science* **2017**,

*356*, 928.

[9] C. L. Degen, F. Reinhard, P. Cappellaro, *Rev. Mod. Phys.* **2017**, *89*, 035002.

[10] J. M. Schloss, J. F. Barry, M. J. Turner, R. L. Walsworth, *Phys. Rev. Appl.* **2018**, *10*, 034044.

[11] P. J. Vetter, A. Marshall, G. T. Genov, T. F. Weiss, N. Striegler, E. F. Großmann, S. Oviedo-Casado, J. Cerrillo, J. Prior, P. Neumann, F. Jelezko, *Phys. Rev. Appl.* **2022**, *17*, 044028.

[12] Z. Yu, Y. Zhu, W. Zhang, K. Jing, S. Wang, C. Chen, Y. Xie, X. Rong, J. Du, *Natl. Sci. Rev.* **2025**, *12*, nwae478.

[13] F. Dolde, H. Fedder, M. W. Doherty, T. Nöbauer, F. Rempp, G. Balasubramanian, T. Wolf, F. Reinhard, L. C. L. Hollenberg, F. Jelezko, J. Wrachtrup, *Nat. Phys.* **2011**, *7*, 459.

[14] T. Iwasaki, W. Naruki, K. Tahara, T. Makino, H. Kato, M. Ogura, D. Takeuchi, S. Yamasaki, M. Hatano, *ACS Nano* **2017**, *11*, 1238.

[15] C.-F. Liu, W.-H. Leong, K. Xia, X. Feng, A. Finkler, A. Denisenko, J. Wrachtrup, Q. Li, R.-B. Liu, *Natl. Sci. Rev.* **2020**, *8*, nwaa194.

[16] N. Wang, G.-Q. Liu, W.-H. Leong, H. Zeng, X. Feng, S.-H. Li, F. Dolde, H. Fedder, J. Wrachtrup, X.-D. Cui, S. Yang, Q. Li, R.-B. Liu, *Phys. Rev. X* **2018**, *8*, 011042.

[17] P. Bhattacharyya, W. Chen, X. Huang, S. Chatterjee, B. Huang, B. Kobrin, Y. Lyu, T. J. Smart, M. Block, E. Wang, Z. Wang, W. Wu, S. Hsieh, H. Ma, S. Mandyam, B. Chen, E. Davis, Z. M. Geballe, C. Zu, V. Struzhkin, R. Jeanloz, J. E. Moore, T. Cui, G. Galli, B. I. Halperin, C. R. Laumann, N. Y. Yao, *Nature* **2024**, *627*, 73.

[18] M. Wang, Y. Wang, Z. Liu, G. Xu, B. Yang, P. Yu, H. Sun, X. Ye, J. Zhou, A. F. Goncharov, Y. Wang, J. Du, *Nat. Commun.* **2024**, *15*, 8843.

[19] D. P. Shelton, W. Cabriales, A. Salamat, *Rev. Sci. Instrum.* **2024**, *95*, 083901.

[20] H.-Y. Liu, W.-Z. Liu, M.-Q. Wang, X.-Y. Ye, P. Yu, H.-Y. Sun, Z.-X. Liu, Z.-X. Liu, J.-W. Zhou, P.-F. Wang, F.-Z. Shi, Y. Wang, *Adv. Quantum Technol.* **2023**, *6*, 2300127.

[21] S.-C. Zhang, Y. Dong, B. Du, H.-B. Lin, S. Li, W. Zhu, G.-Z. Wang, X.-D. Chen, G.-C. Guo, F.-W. Sun, *Rev. Sci. Instrum.* **2021**, *92*, 044904.

[22] F. M. Stürner, A. Brenneis, T. Buck, J. Kassel, R. Rölver, T. Fuchs, A. Savitsky, D. Suter, J. Grimmel, S. Hengesbach, M. Förtsch, K. Nakamura, H. Sumiya, S. Onoda, J. Isoya, F. Jelezko, *Adv. Quantum Technol.* **2021**, *4*, 2000111.

[23] N. Sekiguchi, M. Fushimi, A. Yoshimura, C. Shinei, M. Miyakawa, T. Taniguchi, T. Teraji, H. Abe, S. Onoda, T. Ohshima, M. Hatano, M. Sekino, T. Iwasaki, *Phys. Rev. Appl.* **2024**, *21*, 064010.

[24] J. Shao, Y. Luo, J. Chen, H. Huang, G.-S. Liu, L. Chen, Z. Chen, Y. Chen, *Opt. Express* **2023**, *31*, 14685.

[25] S.-C. Zhang, L. Zhao, R.-J. Qiu, J.-Q. Geng, T. Tian, B.-W. Zhao, Y. Liu, L.-K. Shan, X.-D. Chen, G.-C. Guo, F.-W. Sun, *APL Photon.* **2025**, *10*, 036117.

[26] Q. Liu, S. Nie, X. Peng, Y. Zhu, N. Wang, Y. Hu, X. Luo, C. li, M. Jing, C. Zhang, W. Liu, H. Chen, J. Cheng, Z. Wu, *Adv. Sensor Res.* **2024**, *4*, 2400106.

[27] Y. Hatano, J. Shin, J. Tanigawa, Y. Shigenobu, A. Nakazono, T. Sekiguchi, S. Onoda, T. Ohshima, K. Arai, T. Iwasaki, M. Hatano, *Sci. Rep.* **2022**, *12*, 13991.

[28] K. Kajiyama, M. Haruyama, Y. Hatano, H. Kato, M. Ogura, T. Makino, H. Noguchi, T. Sekiguchi, T. Iwasaki, M. Hatano, *Adv. Quantum Technol.* **2025**, *8*, 2400400.

[29] L. Q. Zhou, R. L. Patel, A. C. Frangeskou, A. Nikitin, B. L. Green, B. G. Breeze, S. Onoda, J. Isoya, G. W. Morley, *Phys. Rev. Appl.* **2021**, *15*, 024015.

[30] A. A. Lanin, I. V. Fedotov, Y. G. Ermakova, D. A. Sidorov-Biryukov, A. B. Fedotov, P. Hemmer, V. V. Belousov, A. M. Zheltikov, *Opt. Lett.* **2016**, *41*, 5563.

[31] F. Perona Martínez, A. C. Nusantara, M. Chipaux, S. K. Padamati, R. Schirhagl, *ACS Sens.* **2020**, *5*, 3862.

[32] D. Duan, V. K. Kavatamane, S. R. Arumugam, G. Rahane, Y.-K. Tzeng, H.-C. Chang, H. Sumiya, S. Onoda, J. Isoya, G. Balasubramanian, *Appl. Phys. Lett.* **2018**, *113*, 041107.

[33] R. L. Patel, L. Q. Zhou, A. C. Frangeskou, G. A. Stimpson, B. G. Breeze, A. Nikitin, M. W. Dale, E. C. Nichols, W. Thornley, B. L. Green, M. E. Newton, A. M. Edmonds, M. L. Markham, D. J. Twitchen, G. W. Morley, *Phys. Rev. Appl.* **2020**, *14*, 044058.

[34] S.-C. Zhang, H.-B. Lin, Y. Dong, B. Du, X.-D. Gao, C. Yu, Z.-H. Feng, X.-D. Chen,

G.-C. Guo, F.-W. Sun, *Photon. Res.* **2022**, *10*, 2191.

[35] D. Kim, M. I. Ibrahim, C. Foy, M. E. Trusheim, R. Han, D. R. Englund, *Nat. Electron.* **2019**, *2*, 284.

[36] Y. Wang, W. Zhang, H. Chai, Z. Zhang, S. Lin, X. Qin, J. Du, *Phys. Rev. Appl.* **2025**, *23,* 034008.

[37] B. Diederich, R. Lachmann, S. Carlstedt, B. Marsikova, H. Wang, X. Uwurukundo, A. S. Mosig, R. Heintzmann, *Nat. Commun.* **2020**, *11*, 5979.

[38] S. Johansson, D. Lönard, I. C. Barbosa, J. Gutsche, J. Witzenrath, A. Widera, *arXiv:* **2025**, 2402.19372.

[39] H. Kumar, S. Dasika, M. Mangat, S. Tallur, K. Saha, *Rev. Sci. Instrum.* **2024**, 95, 075002.

# Supporting Information for
# Fiber-integrated NV Magnetometer with Microcontroller-based Software Lock-in Technique

*Qilong Wu[1†], Xuan-Ming Shen[1†], Yuan Zhang[1, 2 *], Ying-Geng Shan[1], Hui-Hui Yu[1], Jing-Hao Zhang[1], Jiahui Chen[1], Yan Wang[3*], Xun Yang[1], Yong-Zhi Tian[1], Lijun Wang[4], and Chong-Xin Shan[1*]*

[1] Henan Key Laboratory of Diamond Materials and Devices, School of Physics and Laboratory of Zhongyuan Light, Zhengzhou University, Zhengzhou 450052, China

[2] Institute of Quantum Materials and Physics, Henan Academy of Sciences, Zhengzhou 450046, China

[3] School of Electronics and Information, Academy for Quantum Science and Technology, Zhengzhou University of Light Industry, Zhengzhou 450001, China

[4] State Key Laboratory of Luminescence and Applications, Changchun Institute of Optics, Fine Mechanics and Physics, Chinese Academy of Sciences, Changchun 130033, China

*Email: yzhuaudipc@zzu.edu.cn; ywang@zzuli.edu.cn; cxshan@zzu.edu.cn

In this supporting information, we provide further information to complement the discussion in the main text.

**Section S1: Details of Fiber-integrated NV Magnetometer**

In this section, we provide details on the fiber-integrated NV magnetometer. The electronic module is shown in Figure S1(a), which consists of a microcontroller (STM32F103ZET6), a microwave source chip (ADF4351), a constant current source (CCS KW-VCCS1000), a digital attenuator (HMC624A), an amplifier (KDT0308B-004), and a power supply (XL4005). The connection of the electronics is shown in Figure S1(b). The microcontroller communicates with the microwave chip and the attenuator chip through Serial Peripheral Interface (SPI) protocol, and configures their registers to realize the specific functions. In this way, the microwave chip can output the radiation in the frequency range from 35 MHz to 4400 MHz, with four selectable power levels (-4 dBm, -1 dBm, 2 dBm, 5 dBm). The attenuator can reduce the microwave radiation in the range from -30 dB to 0 dB, and the amplifier can amplify the radiation by up to 40 dB. The microcontroller uses an integrated digital-to-analog converter (DAC) to output a voltage from 0 V to 3.3 V. This voltage regulates the constant current source to output the current from 0 mA to 330 mA, which powers the laser diode. The microcontroller uses also an integrated analog-to-digital converter (ADC) to sample the output voltage from the photodetector. In addition, the minimal system with the microcontroller chip features a USB-to-serial chip (CH340), and thus can communicate with the computer through a serial-port based protocol.

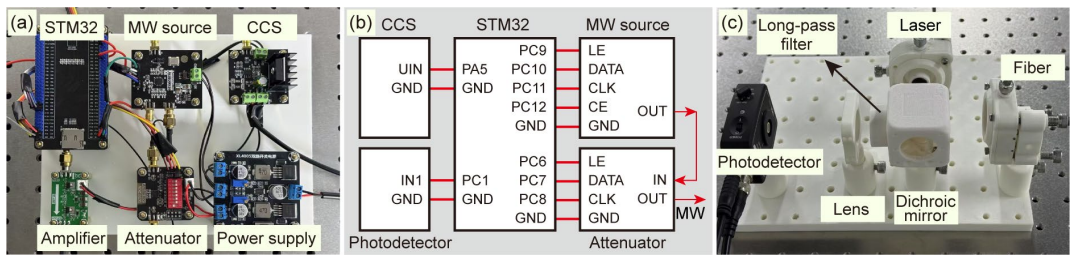

**Figure S1. Photograph of the magnetometer**. (a) Electronic module with the parts marked. (b) Circuit connections of the electronics. (c) Optical module with the parts marked. All components are mounted on the 3D printed optical boards.

The optical module is shown in Figure S1(c). All mirrors are mounted on 3D-printed components. Although commercially available optical components exist, they are typically expensive and offer limited flexibility for miniaturization. In contrast, 3D printing technique provides a cost-effective and versatile alternative, enabling rapid prototyping and iterative optimization.

**Section S2: Sensing Head Fabrication**

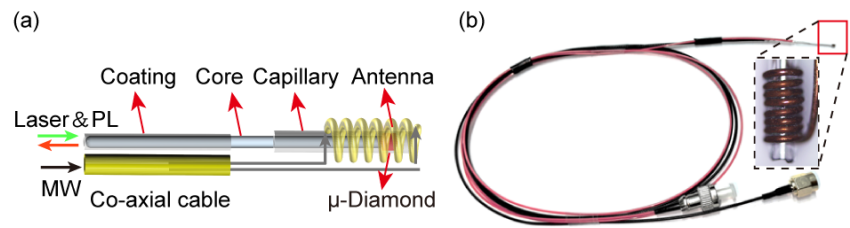

**Figure S2. Fabrication of the fiber-integrated sensing head.** (a) Schematic of the sensing head with the main components marked. (b) Photo of the fabricated sensing head and the zoom-in of the core region.

In this section, we describe the fabrication process of the fiber-integrated sensing head. The sensing head consists of a multi-mode optical fiber with a core diameter 105 μm, a copper coil with a diameter 150 μm, and a microdiamond of 40 μm size with 3.5 ppm NV$^-$ center concentration from Adámas [Figure S2(a)]. The fabrication involves four main steps. First, the outer protective layer and coating of the fiber are removed, and then the fiber is cleaved to produce a flat and smooth facet. Second, the microdiamond particles are scattered on a glass slide, and a single microdiamond is then picked up with a fiber facet with UV-curable adhesive while being monitored under a microscope. The fiber head is then exposed to UV light for 5 minutes to fix the microdiamond. Third, to protect the sensing head, the microdiamond-integrated fiber is inserted into a capillary tube with 300 μm internal diameter, around which a copper wire is wound several turns to serve as a microwave antenna, and is then fixed with UV adhesive. Finally, the microwire is soldered to the core and the cover layer of a co-axial cable, and the fiber and cable are connected to the optical-fiber coupler and the amplifier. Figure S2(b) shows the photograph of the fabricated sensing head, along with a zoomed-in view of the antenna.

**Section S3: Software Lock-in Technique Code**

```
(1)  for (i = 0; i<4; i++){
         command1[i] = comm_str[i+5];
     }
     if (strcmp(command1,"DEVI") ==0 ){
         for (i=0; i< rec_num - 12; i++)
         {
             argument0[i] = comm_str[i+10];
         }
         odmr_fm_freq_devi = atof(argument0);
     }
(2)  if (strcmp(command1,"STAR") ==0 ){
         odmr_freq_step = odmr_freq_width/odmr_freq_nums;
         odmr_freq_start = odmr_freq_center - odmr_freq_width/2;

(3)      for (i =0; i< odmr_freq_nums; i++){
             x_vec[i] = odmr_freq_start + i*odmr_freq_step;
             odmr_fm_freq_up = x_vec[i] + odmr_fm_freq_devi*0.5;
             odmr_fm_freq_down = x_vec[i] - odmr_fm_freq_devi*0.5;
             ADF4351WriteFreq(odmr_fm_freq_up);
             delay_us(1000);
             adc_on =0;
             for (j =0; j< odmr_samp_nums; j++){
                 adc_on += (float) ADC_ConvertedValue/4096*3.3;
                 delay_us(10);
             }
             odmr_fm_up_adc = adc_on/odmr_samp_nums;
             ADF4351WriteFreq(odmr_fm_freq_down);
(4)          delay_us(1000);
             adc_off =0;
             for (j =0; j< odmr_samp_nums; j++){
                 adc_off += (float) ADC_ConvertedValue/4096*3.3;
                 delay_us(10);
             }
             odmr_fm_down_adc = adc_off/odmr_samp_nums;
             y_vec[i] = odmr_fm_up_adc-odmr_fm_down_adc;
         }
     }
```

**Figure S3. Partial code for the software lock-in technique**

The part of the code that implements the microcontroller-based software lock-in technique is shown in Figure S3. The first code segment is responsible for setting the FM modulation depth by parsing the deviation value from the received command string. Based on the center frequency, the frequency scanning width, and the number of sampling points, the second segment calculates the scanning frequency step and the starting frequency. The core of the program, implemented in the third and fourth code segments, realizes the software lock-in technique. Starting from the first frequency point, the upper and lower frequencies are calculated by adding and subtracting half of the modulation depth, respectively. These two frequencies are sequentially sent to the microwave chip with a short delay to allow stabilization. At each frequency, the photodetector output voltages are sampled by the ADC, then summed and averaged. Finally, the difference between the average voltage at the upper and lower frequencies is computed to mimic a low-pass filter. By varying the microwave carrier frequency, the frequency-modulated optically detected magnetic resonance (FM ODMR) is obtained.

**Section S4: Comparison with Commercial Equipment**

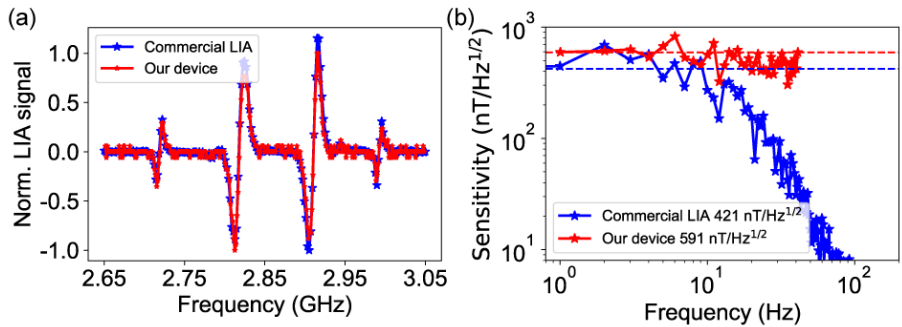

**Figure S4. Comparison of FM ODMR (a) and magnetic field sensitivity (b) with the commercial LIA (blue) and our device (red), respectively.**

In this section, we compare our setup with a commercial setup, which is based on an FPGA-based commercial LIA (CIQTEK, LIA001M) and a commercial microwave source (RIGOL, DSG 863A). LIA001M is configured to output a square reference signal with a frequency of 2033 Hz, and to demodulate the signal with a filter of 10 Hz bandwidth, while the DSG 836A is set to the maximal modulation depth of 1 MHz. For both setups, the microwave power after amplification is 25 dBm. Figure S4(a) shows that the FM ODMR spectra from both setups are nearly identical. We then set the microwave frequency to the third zero-crossing point. The measured sensitivity below 10 Hz is around 421 $\mathrm{nT}/\sqrt{\mathrm{Hz}}$ for the commercial setup and 591 $\mathrm{nT}/\sqrt{\mathrm{Hz}}$ for our setup [Figure S4(b)]. The former is only 1.3 times better than the latter, but our device is more cost-effective and portable, demonstrating the potential in industrial applications.